# Valence-Electron Transfer and a Metal-Insulator Transition in a Strongly Correlated Perovskite Oxide

A. P. Ramirez[1], G. Lawes[2], D. Li[3], and M. A. Subramanian[3]

[1]*Bell Laboratories, Lucent Technologies, 600 Mountain Avenue, Murray Hill, NJ 07974*
[2]*Los Alamos National Laboratory, Los Alamos, NM 87545*
[3]*Dupont Central Research and Development, Experimental Station, Wilmington, Delaware, 19880-0328*

We present transport and thermal data for the quadruple-perovskites $MCu_3(Ti_{1-x}Ru_x)_4O_{12}$, where $0 < x < 1$. A metal-insulator transition (MIT) occurs for Ru concentrations $x \sim 0.75$. At the same time, the $Cu^{2+}$ antiferromagnetic state is destroyed and it's magnetic entropy suppressed by Ru on a 1:1 basis. This implies that each Ru transfers an electron to a Cu ion and thus the MIT correlates with filling the Cu 3*d* shell. The Cu spin entropy in this strongly correlated electron material provides a unique probe among MIT systems.

The interplay between different ground states is a central theme in condensed matter physics. One of the more important examples of such interplay is the transition between itinerant and localized electrons, the metal-insulator transition (MIT). MITs have been studied in a wide variety of ionic solids but are realized with great diversity in compounds based on Ru-O octahedra. Ruthenates display ground states that are ferromagnetic (FM), antiferromagnetic (AF) [1, 2], superconducting (SC) [3, 4], Fermi-liquid-like (FL) [5], and non-Fermi-liquid-like [6]. These materials can be easily tuned with pressure, magnetic field and disorder and several different types of MITs have been observed. This large variety of behaviors is an indication of the high-sensitivity of Ru to its local environment [7]. Surprisingly, the ground state in a particular compound can be insensitive to the Ru *formal* valence – for instance, $SrRuO_3$ and $Sr_2RuO_4$ are ferromagnetic and (p-wave) superconducting, respectively, despite Ru having a formal valence of +4 in each compound. It is clearly of great importance to understand the energetics of valence tuning in Ru-O based systems, as the microscopic basis for the unusual variety of ruthenate ground states.

Here we study the transport and thermodynamic behavior of a perovskite-based compound, $MCu_3(Ti_{1-x}Ru_x)_4O_{12}$, where M = Na, La, or Ca, and $0 \leq x \leq 1$. The structure of $AB_3C_4O_{12}$ can be thought of as a perovskite with a BCC superstructure imposed on it by the A-ion, the "quadruple perovskite" structure, as shown in figure 3. Note that the $TiO_6$ or $RuO_6$ octahedra are surrounded by a network of interleaving chains of $CuO_4$ plaquettes. The x = 0 compound, $CaCu_3Ti_4O_{12}$ has attracted much interest due to it's large high-temperature dielectric constant [8-10], originating most likely in the $TiO_6$ subsystem [11, 12]. However, the behavior of the $Cu^{2+}$ network seems simple – the associated s = ½ moments undergo AF order at $T_N$ = 25K into a collinear state [13], with an associated lambda-type anomaly in the specific heat [14]. The Ru end-member $CaCu_3Ru_4O_{12}$ has also been synthesized and is known to be a Fermi-liquid down to 2K [5, 15]. The complex structure of $AB_3C_4O_{12}$ in particular, with its three distinct cation sites, allows for great flexibility in chemical substitution [5]. In particular, we find that the collective behavior of the localized $Cu^{2+}$ moment allows a precise measure of the degree of charge transfer, and hence valence tuning, between Cu and Ru ions. We find that an MIT does not occur until 75% of the Ti are replaced by Ru. At the same time, the localized $Cu^{2+}$ entropy decreases linearly with increasing Ru content, vanishing at the same concentration where metallic behavior is seen.

Samples of $MCu_3X_4O_{12}$ (M = Na, Ca and La, X = Ru, Ti) were prepared by standard solid reaction using high purity carbonates (Na and Ca) and oxides (La, Ru, Ti, and Cu). The starting materials were mixed thoroughly and heated to 800-950C for 24-36 hours with intermediate grindings. X-ray diffraction patterns could be indexed on a cubic unit cell and the observed unit cell parameters for the x = 0 end member is in agreement with the published values. Measurements of the resistance were made using a standard four-wire method. Measurements of the susceptibility were made using a commercial Superconducting Quantum Interference Device (SQUID) magnetometer. Measurements of the heat capacity were made using both a commercial calorimeter system employing the relaxation method (Quantum Design) for measurements above 2K, and in a top-loading $^3He$-$^4He$ dilution refrigerator using a semiadiabatic technique for measurements below 2K.

We first discuss the properties of compounds with varying A-site occupancy. In fig. 1 are shown normalized resistivity data for $MCu_3Ru_4O_{12}$ where M = Na, Ca, and La. We note that all of these compounds display metallic behavior varying as $\rho = \rho_0 + AT^2$, with a residual resistivity $\rho_0$ that scales with the A-site occupant: $\rho_0(Na) = 1.35\times10^{-5}$ Ω-cm, $\rho_0(Ca) = 7.27\times10^{-5}$ Ω-cm, $\rho_0(La) = 3.76\times10^{-4}$ Ω-cm. Table 1 lists the lattice constants ($a_0$) for this series of compounds.

The upper inset of fig.1 shows that *A* increases monotonically with $a_0$ as expected from simple band-narrowing related to the larger cation on the A-site. Band-narrowing is also reflected in the dependence of the linear term, $\gamma$, of the Fermi-liquid specific heat $C(T) = \gamma T$, as shown in fig. 1 (inset) for different A-cations.

We note that the changing *A*-cation size is accompanied by a changing formal valence count. Na, Ca, and La normally assume the valence 1+, 2+, and 3+, respectively. For n-type conduction, the trends in *A* and $\gamma$ are consistent with a larger formal valence on the A-site translating into a larger carrier density. This shows that, for these Ru compounds, the bandwidth is large enough that a change of electron count on the Ru-Cu-O sublattice of 2/7 is comparable to the lattice-constant change for modifying the density of states at the Fermi level. This can be contrasted with the situation for $La_{1-x}Sr_xTiO_3$, where an MIT is achieved through band-filling tuning due to a near unity change of valence on the B-cation site [16]. We will see below that in the present system, the role of the A-cation and associated lattice constant changes are secondary to the effects of Cu-Ru interactions.

We now discuss the behavior of the transition from insulating $MCu_3Ti_4O_{12}$ to the metallic $MCu_3Ru_4O_{12}$. In fig. 2 are shown resistivity versus temperature data for members of the series $LaCu_3(Ti_{1-x}Ru_x)_4O_{12}$. We find that for $x \leq 0.75$, the behavior is semiconducting with a transport gap, assuming $\rho(T) = \rho_0 e^{\Delta/T}$, of $\Delta \sim 0.2$ eV. For $x > 0.75$, the behavior is metallic over the entire temperature range. For $x = 0.75$, the high-temperature behavior is metallic, $d\rho/dT > 0$, while below 25K, $d\rho/dT < 0$, showing that the MIT occurs at a concentration very close to 75% Ru.

The susceptibility, $\chi(T)$, for different Ru concentrations is shown in fig. 3. We see for $x = 0$ and M = Ca, $\chi(T)$ behaves like an antiferromagnet, reflecting the fluctuation and ordering of $Cu^{2+}$ moment [14]. As Ru is added, the signature from $Cu^{2+}$ moments above $T_{Neel}(x=0) = 25K$ decreases roughly in proportion to the Ru concentration. Finally, at x = 0.90 the susceptibility approaches a constant Pauli-like value of ~$8.3 \times 10^{-4}$ emu/moleRu. It is difficult to estimate the effect of Ru concentration on the Cu susceptibility. However if we assume the contribution from Ru spin fluctuations (see below) is negligible at the lowest temperatures where Curie-Weiss behavior is observed, we find for $x \leq 0.75$, that the effective density of $Cu^{2+}$ moments, n(x) scales with x as shown in fig. 4, inset. This behavior and it's relation to $\gamma$ will be discussed below.

From a comparison of the temperature-independent value of $\chi(T)$ and the low-temperature value for $\gamma$, we can estimate the importance of spin fluctuations coming from the Ru electrons in the x = 0.9 sample. After subtracting a small Curie term, the temperature-independent susceptibility is $\chi_{Pauli} = 7.7 \times 10^{-4}$ emu/mole-Ru. From fig. 4 and fig. 1 (inset) we find $\gamma = 0.033$ J/moleK$^2$. We thus get a Wilson ratio $R_W = \chi_{Pauli}(\pi k_B)^2/3\gamma\mu_B^2 = 1.6$, which is similar in magnitude to many heavy fermion systems with moderate effective mass [17]. The magnitude of $R_W$ indicates that the mass enhancement in $MCu_3Ru_4O_{12}$ arises primarily from spin fluctuations strongly coupled to the charge carriers.

The vanishing of $Cu^{2+}$ moments in response to increasing Ru concentration is most clearly seen in the specific heat, shown in fig. 4. For x = 0, M = Ca, the specific heat displays a $\lambda$-type anomaly peaking at the Neel temperature [14]. For x = 0.25 and M = Ca or La, the specific heat associated with $Cu^{2+}$ has decreased significantly. This behavior continues with increasing Ru content until at x = 0.75, the local moment $Cu^{2+}$ entropy has disappeared, yielding to a Fermi-liquid signature, consistent with the $\rho(T)$ and $\chi(T)$ data. In the inset of figure 4 we show the magnetic entropy, $\Delta S(x)$, with an approximation to the lattice contribution subtracted. We see that the behavior of $\Delta S(x)$ is nearly linear with respect to Ru concentration. Thus we have the empirical result that for every Ru, one Cu moment is removed.

As shown above, the simultaneous vanishing of $Cu^{2+}$ moment and the appearance of a metallic state are intimately related. We note that the long range ordering of $Cu^{2+}$ moments is very sensitive to Ru incorporation and several other aspects should be noted. First, Ru occupies a distinctly different lattice site from Cu, and single-crystal X-ray and neutron powder structure refinements show no evidence of intersite mixing, a possible source of disorder, in these phases [5, 8]. Second, $Ru^{4+}$ should substitute for $Ti^{4+}$ without lattice distortion due to their similarity in ionic size and charge. Third, long range order (LRO) is destroyed at very small (< 25%) concentrations of Ru. This suggests that the mechanism for destroying LRO is a long-ranged interaction, and not due to a simple dilution of quenched moments. One possible scenario is that, unlike $Cu^{2+}$ and $Ti^{4+}$ valence states, the energy separation between the $Cu^{2+}$ 3*d* and $Ru^{4+}$ 4*d* bands is smaller than the corresponding bandwidths, leading to strong hybridization among the valence electron states. If this were a static effect, and the $Cu^{2+}$ moment reduced by an effective charge-transfer leading to a $Cu^{1+}$ state, then Ru substitution should have a similar effect on LRO as Cu-ion dilution with a nonmagnetic 2+ species. Such a dilution would have a weaker effect on LRO – $T_{Neel}$ would be suppressed linearly with x but the specific heat anomaly would remain sharp as long as $T_{Neel} > 0$. Instead, the anomaly is significantly rounded even at x = 0.25 indicating LRO is destroyed at a much lower value than implied by a simple dilution mechanism. This suggests that while each Ru ion quenches a net single $Cu^{2+}$ moment, it does so by forming a fluctuating valence bond with its eight nearest Cu neighbors.

It is instructive to compare the present dilution study with those of other Ru oxides. Studies such as ours, where the Ru site is held at the same nominal charge value, fall into two categories. The first type of study involves isovalent cation substitution on the non-Ru site, e.g $Sr^{2+}$ for $Ca^{2+}$.

For the 3D structure $Sr_{1-y}Ca_yRuO_3$, a FM to AF transition is induced for y ranging between 0 and 1 but the material remains metallic over the entire dilution range [18]. For the 2D case, $Sr_{2-y}Ca_yRuO_4$, the material transforms from a superconductor for y = 0 to an AF insulator for y = 1, with an MIT critical concentration near y = 0.2 [19]. However, the MIT is accompanied by a series of structural transitions involving rotation and tilt of the $Ru-O_6$ octahedra [20]. It is thought that the MIT is thus driven by the resulting changes in orbital orientation, and hence bandwidth and hybridization modification.

The second type of study is most similar to the present one. Here the Ru-site itself is diluted with an isovalent ion. In $SrTi_{1-x}Ru_xO_3$, for example, the critical MIT concentration is x = 0.35, consistent with a simple percolative mechanism for conduction [21]. Given the similarity of structure between the perovskite $SrTiO_3$ and the present quadruple-perovskite structure the occurrence of an MIT at x = 0.75 is significant. A clue to this difference in behavior comes from the structure. $CaCu_3Ti_4O_{12}$ is a true quaternary, and in particular, the structure supports a 3*d* ion, $Cu^{2+}$ in the A cation site. A close relative of the present compound is found in $RuSr_2GdCu_2O_8$ which is thought to exhibit simultaneous FM and superconducting states [22]. Here, however, the quasi-2D structure avoids the steric constraint imposed by the tolerance factor, allowing Cu and Ru to coexist in the same structure.

Since $Cu^{3+}$ ($d^8$) is magnetic, and the Cu moment vanishes linearly with Ru incorporation, it is clear that each Ru transfers an electron to a Cu ion, thus filling the Cu 3*d*-shell, and not a hole. Since the ratio of Cu to Ru sites is 3:4, this charge transfer process is exhausted at roughly x = 0.75. For higher concentrations, the Ru valence electrons are itinerant. Thus, the present MIT is a variation on a Mott transition with the Ru concentration tuning the carrier density. The unusual aspect of $MCu_3(Ru,Ti)_4O_{12}$ is that the tuning is strongly controlled by the presence of Cu as evidenced from the x-dependence of the AF entropy. Thus, this novel type of MIT is a result of the unusual quadruple perovskite structure, which allows for a distinct $Cu^{2+}$ sublattice in close proximity to the Ru-Ti sublattice.

We would like to acknowledge useful discussions with P. B. Littlewood and C. M. Varma and thank T. G. Calvarese (DuPont) for technical assistance. We also acknowledge the support of the Los Alamos LDRD program for the specific heat measurements below 2K.

**Figure Captions**

Fig. 1. Temperature dependence of resistivity for the metallic end members with different A-site cations. The lower inset shows the low-temperature specific heat. The upper inset show the values of *A* and $\gamma$ that parameterize these data as a function of lattice constant.

Fig. 2 Resistivity for Ru-susbstituted compounds. For $x \leq 0.75$, the ground state is insulating.

Fig. 3. The dc-susceptibility for compounds with varying Ru concentrations. The structure of $ACu_3(Ti,Ru)_4O_{12}$ is shown in the inset.

Fig. 4 Specific heat of the compounds $MCu_3(Ti_{1-x}Ru_x)_4O_{12}$ for various M and x values. The inset shows the magnetic entropy of $Cu^{2+}$ spins, $\Delta S(x)$, and an estimate of the density of free spins, n(x) obtained from the dc-susceptbility. The error bar on $\Delta S$ is an estimate of the uncertainty associated with the lattice subtraction (see text).

**Table 1.**

| | a (Å) | Literature value (Å)* |
|---|---|---|
| $CaCu_3Ru_4O_{12}$ | 7.452(3) | 7.421 |
| $LaCu_3Ru_4O_{12}$ | 7.485(2) | 7.479 |
| $NaCu_3Ru_4O_{12}$ | 7.382(1) | 7.390 |

*J Solid State Chem **33**, 257-261 (1980)

Lattice parameters of $LaCu_3Ru_{4-x}Ti_xO_{12}$:

| | x | a (Å) | 95% confident |
|---|---|---|---|
| $LaCu_3Ru_4O_{12}$ | 0 | 7.4851 | 0.0042 |
| $LaCu_3Ru_4O_{12}$ | 0 | 7.479 | - |
| $LaCu_3Ru_{3.6}Ti_{0.4}O_{12}$ | 0.4 | 7.4799 | 0.0006 |
| $LaCu_3Ru_3TiO_{12}$ | 1 | 7.4805 | 0.0022 |
| $LaCu_3Ru_2Ti_2O_{12}$ | 2 | 7.4547 | 0.0034 |
| $LaCu_3RuTi_3O_{12}$ | 3 | 7.4365 | 0.0048 |
| $La_{2/3}Cu_3Ti_4O_{12}$ | 4 | 7.4271 | 0.0061 |

Lattice parameters of $CaCu_3Ru_{4-x}Ti_xO_{12}$:

| | x | a (Å) | 95% confident |
|---|---|---|---|
| $CaCu_3Ru_4O_{12}$ | 0 | 7.4516 | 0.0073 |
| $CaCu_3Ru_4O_{12}$ | 0 | 7.421 | - |
| $CaCu_3Ru_{3.6}Ti_{0.4}O_{12}$ | 0.4 | 7.4052 | 0.0037 |
| $CaCu_3Ru_3TiO_{12}$ | 1 | 7.4124 | 0.0136 |
| $CaCu_3Ru_2Ti_2O_{12}$ | 2 | 7.3905 | 0.0023 |
| $CaCu_3RuTi_3O_{12}$ | 3 | 7.3929 | 0.0022 |
| $CaCu_3Ti_4O_{12}$ | 4 | 7.3908 | 0.0012 |

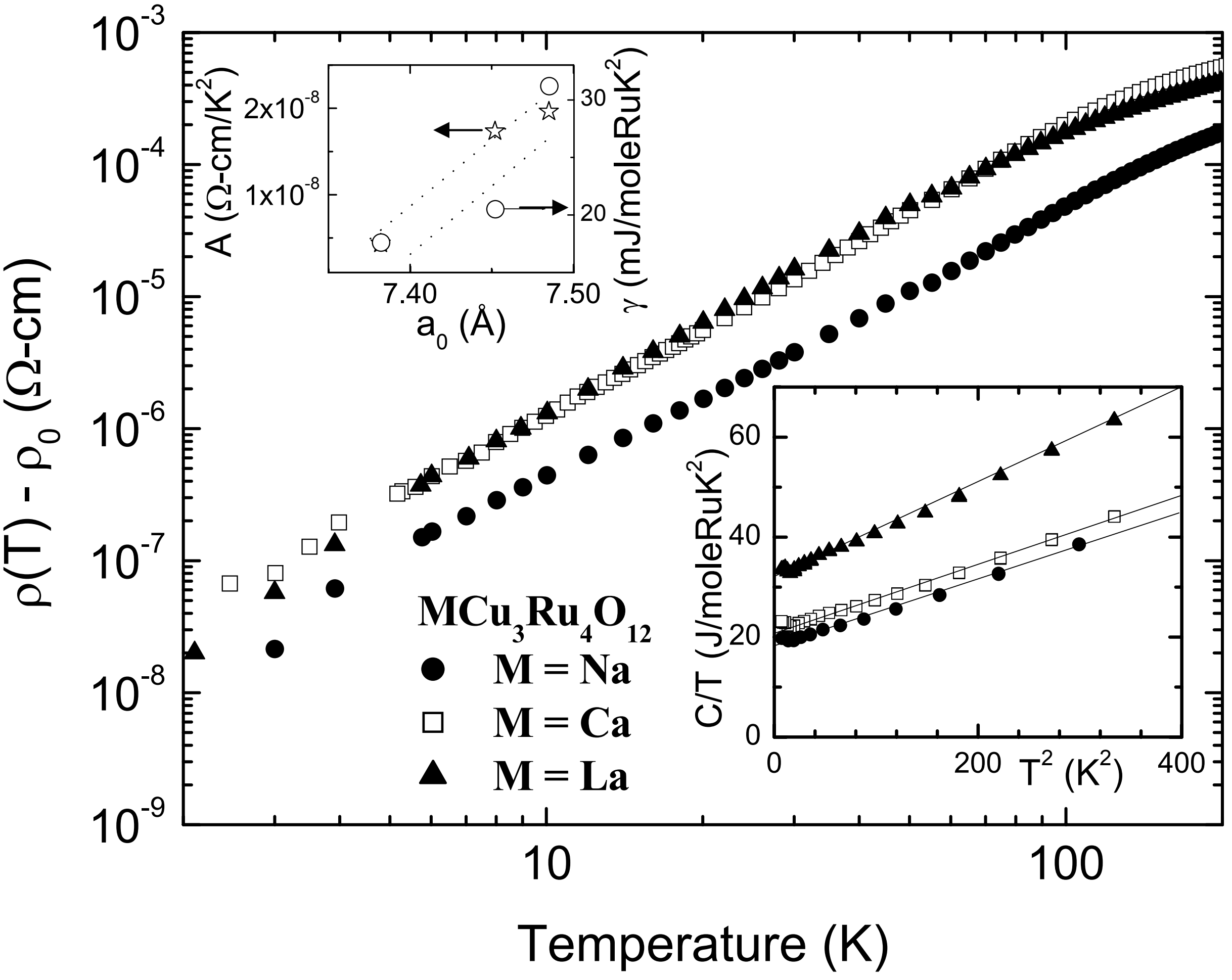

Fig. 1

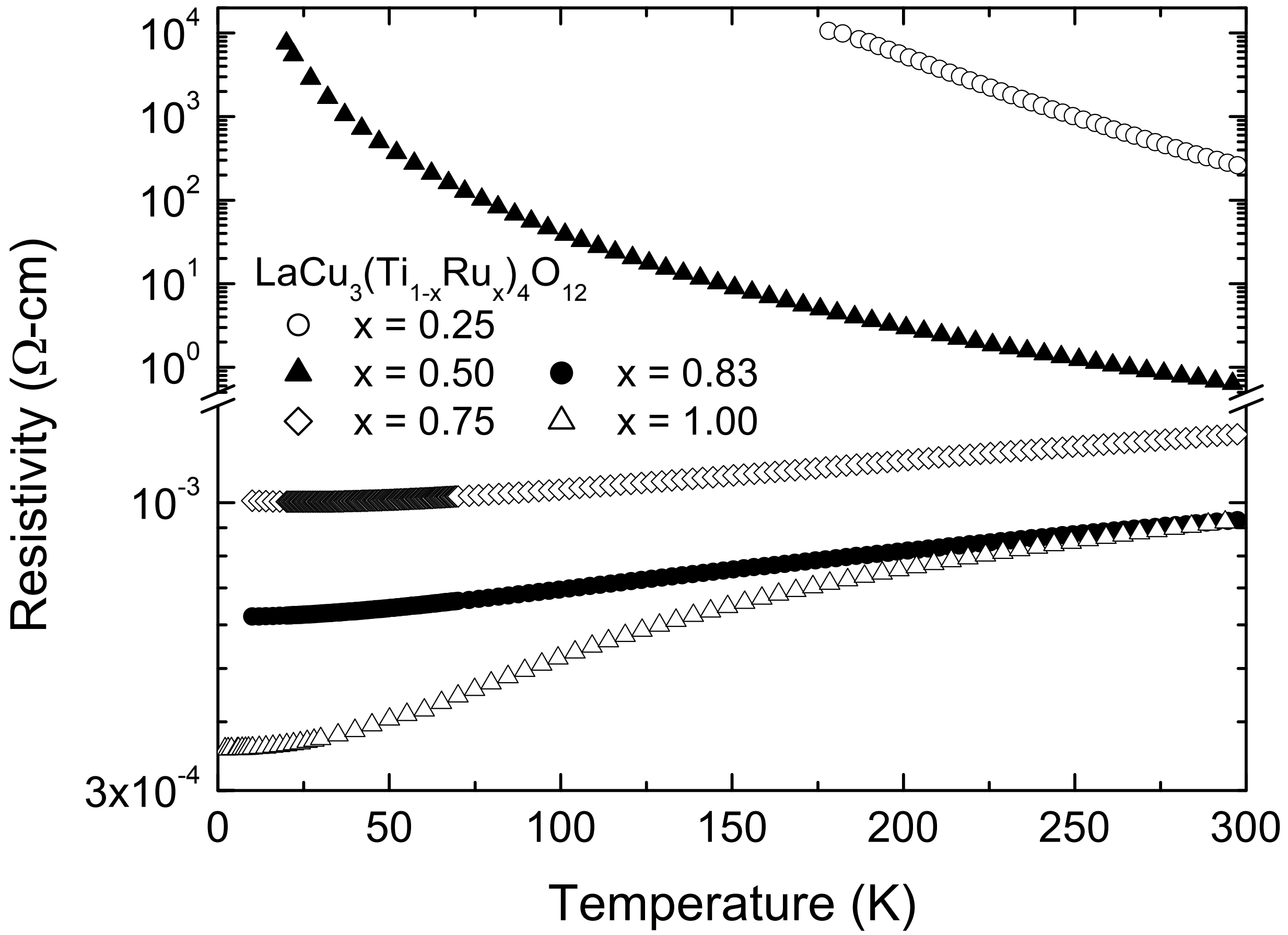

Fig. 2

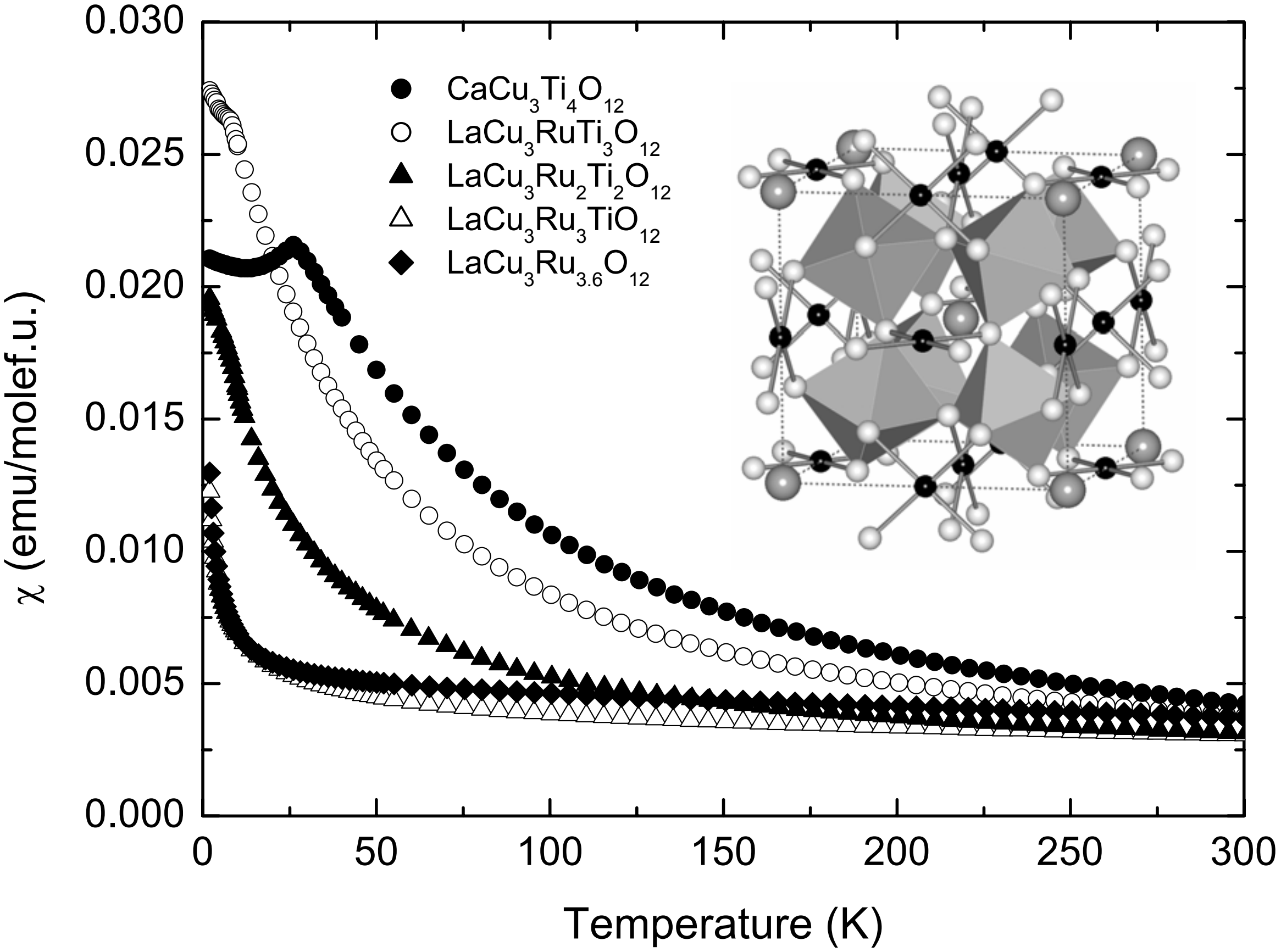

Fig. 3

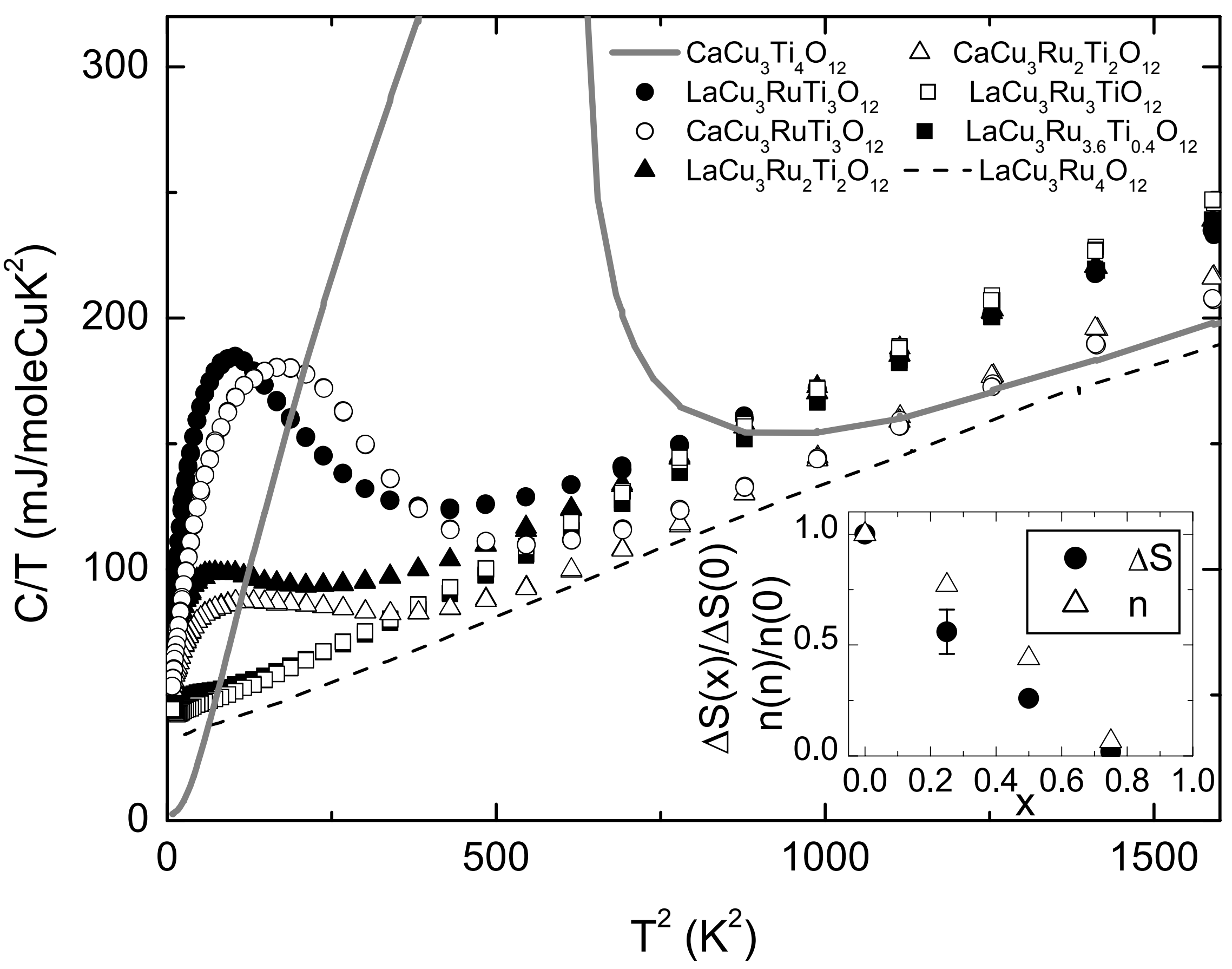

Fig. 4